\documentclass{ws-procs9x6}

\begin{document}

\title{Towards exploring $U_{\lowercase{e}3}$
\footnote{\uppercase{T}his work is supported in part by
the \uppercase{G}rants-in-\uppercase{A}id for \uppercase{S}cientific
\uppercase{R}esearch
in \uppercase{P}riority \uppercase{A}reas \uppercase{N}o. 12047222
and \uppercase{N}o. 13640295,
\uppercase{J}apan \uppercase{M}inistry
of \uppercase{E}ducation, \uppercase{C}ulture, \uppercase{S}ports,
\uppercase{S}cience, and \uppercase{T}echnology.}}

\author{Osamu Yasuda}

\address{Department of Physics, Tokyo Metropolitan University,\\
1-1 Minami-Osawa,
Hachioji, Tokyo 192-0397, Japan\\
E-mail: yasuda@phys.metro-u.ac.jp}


\maketitle

\abstracts{
It is emphasized that resolution of the $\theta_{23}$ ambiguity
is important for determination of $\theta_{13}$ if
$\sin^22\theta_{23}<1$, and resolution of the sgn($\Delta m^2_{31}$)
ambiguity is important for determination of the CP phase $\delta$.
I discuss the prospects of resolution of the $\theta_{23}$
ambiguity etc. in the future long baseline experiment
after the JPARC
experiment measures the oscillation probabilities
$P(\nu_\mu\rightarrow\nu_e)$ and
$P(\bar{\nu}_\mu\rightarrow\bar{\nu}_e)$ at $|\Delta m^2_{31}|L/4E=\pi/2$.
}

\section{Introduction}
From the recent experiments on atmospheric and
solar, and reactor neutrinos, we now know approximately
the values of the mixing angles and the mass squared differences
of the atmospheric and solar neutrino oscillations:
$(\sin^22\theta_{12}, \Delta m^2_{21})\simeq
(0.8, 7\times10^{-5}{\rm eV}^2)$ for the solar neutrino
and $(\sin^22\theta_{23}, |\Delta m^2_{31}|)\simeq
(1.0, 2\times10^{-3}{\rm eV}^2)$ for the atmospheric neutrino.
In the three flavor framework of neutrino oscillations,
the quantities which are still unknown to date are
the third mixing angle $\theta_{13}$, the sign
of the mass squared difference $\Delta m^2_{31}$ 
of the atmospheric neutrino oscillation, and
the CP phase $\delta$.
It is expected that these three quantities will be determined
by long baseline experiments in the future.

It has been
known that even if the values of the oscillation
probabilities $P(\nu_\mu \rightarrow \nu_e)$ and
$P(\bar{\nu}_\mu \rightarrow \bar{\nu}_e)$ are exactly given
we cannot determine uniquely the values of the
oscillation parameters due to parameter degeneracies.
There are three kinds of parameter degeneracies:
the intrinsic $(\theta_{13}, \delta)$ 
degeneracy,
the degeneracy of
$\Delta m^2_{31}\leftrightarrow-\Delta m^2_{31}$,
and the degeneracy of
$\theta_{23}\leftrightarrow\pi/2
-\theta_{23}$.
Each degeneracy gives a twofold solution, so
in total we have an eightfold solution if
all the degeneracies are exact.  In this case
prediction for physics is the same for all the
degenerated solutions and there is no problem.
However, at least two out of the three degeneracies
are lifted slightly in long baseline experiments,
and there are in general eight
different solutions.
When we try to determine the oscillation parameters,
ambiguities arise because the values of the oscillation parameters are
slightly different for each solution.
In particular, this causes a serious problem in measurement of
CP violation, which is expected to be small effect
in the long baseline experiments, and we could mistake
a fake effect due to the ambiguities for
nonvanishing CP violation if we do not treat the ambiguities carefully.

In this talk, assuming the JPARC experiment measures
the oscillation probabilities $P(\nu_\mu\rightarrow\nu_e)$ and
$P(\bar{\nu}_\mu\rightarrow\bar{\nu}_e)$ at the oscillation
maximum (i.e., with $|\Delta m^2_{31}|L/4E=\pi/2$),
I will discuss the possibilities for an experiment following
JPARC to determine $|U_{e3}|$ and arg($U_{e3}$).
The details of the present discussions and references
are found in Ref.~\cite{Yasuda:2004gu}.

\section{$|U_{e3}|$}
In this section, assuming that the JPARC experiment measures
$P(\nu_\mu \to \nu_e)$ and
$P(\bar\nu_\mu \to \bar\nu_e)$
at the oscillation maximum $\Delta\equiv|\Delta m^2_{31}|L/4E=\pi/2$,
I will discuss how the third measurement after JPARC
can resolve the ambiguities by using the plot in the
($\sin^22\theta_{13}$, $1/s^2_{23}$) plane.

First of all, let me consider the case where experiments are
done at the oscillation maximum, i.e., when the neutrino
energy$E$ satisfies $\Delta=\pi/2$.
In this case, the trajectory of $P(\nu_\mu \to \nu_e)=P$,
$P(\bar\nu_\mu \to \bar\nu_e)=\bar{P}$
becomes a straight line
in the ($X\equiv\sin^22\theta_{13}$, $Y\equiv1/s^2_{23}$) plane
and is given by
\begin{eqnarray}
Y=\frac{f+\bar{f}}{P/f+\bar{P}/\bar{f}-C(1/f+1/\bar{f})}
\left(X-\frac{C}{f\bar{f}}\right)
\label{omn}
\end{eqnarray}
for the normal hierarchy, and
\begin{eqnarray}
Y=\frac{f+\bar{f}}{P/\bar{f}+\bar{P}/f-C(1/f+1/\bar{f})}
\left(X-\frac{C}{f\bar{f}}\right)
\label{omi}
\end{eqnarray}
for the inverted hierarchy, where
\begin{eqnarray}
\left\{ \begin{array}{c}
f\\ \bar{f} \end{array}\right\}
\equiv \pm\frac{\cos(AL/2)}{1\mp AL/\pi},~~
C&\equiv&\left(\frac{\Delta m^2_{21}}{\Delta m^2_{31}}
\right)^2\left[\frac{\sin(AL/2)}{AL/2\Delta}
\right]^2\sin^22\theta_{12},
\nonumber
\end{eqnarray}
and $A\equiv\sqrt{2}G_FN_e$ is the matter effect.
Since Eqs. (\ref{omn}) and (\ref{omi}) are linear in $X$,
there is only one solution between them and $Y$=const.
Thus the ambiguity due to the intrinsic degeneracy is solved
by performing experiments at the oscillation maximum, although
it is then transformed into
another ambiguity due to the $\delta\leftrightarrow\pi-\delta$ degeneracy.
\begin{figure}
\vglue -0.4cm
\hglue -0.4cm
\includegraphics[scale=0.4]{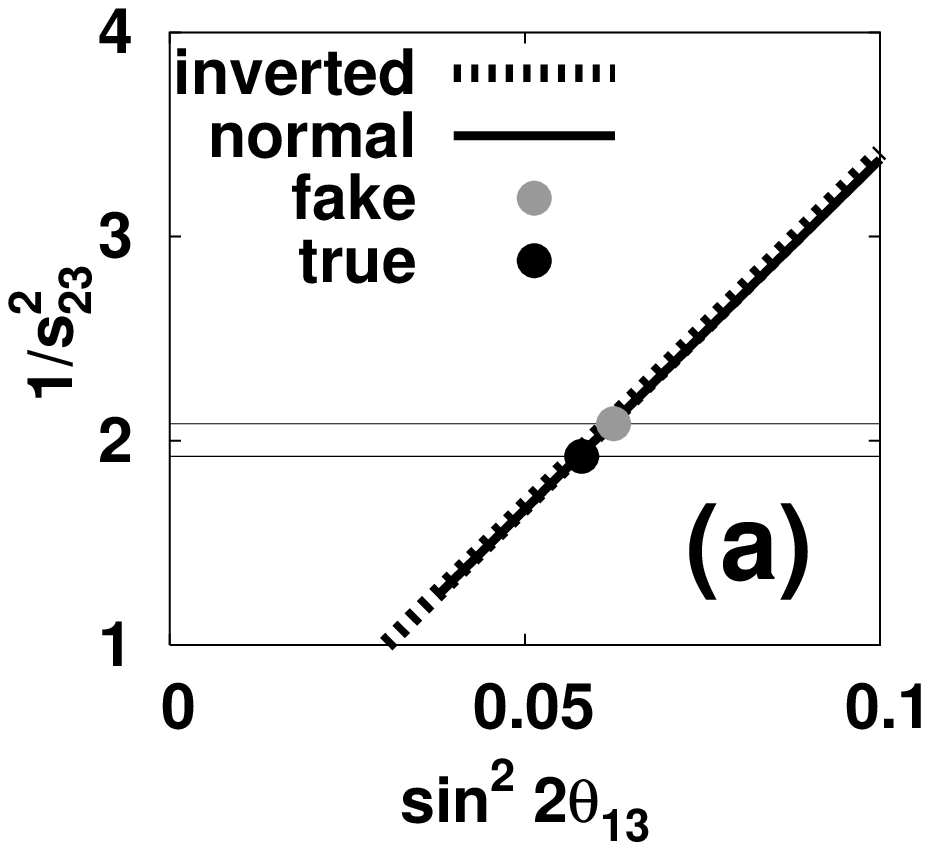}
\hglue -0.3cm
\includegraphics[scale=0.4]{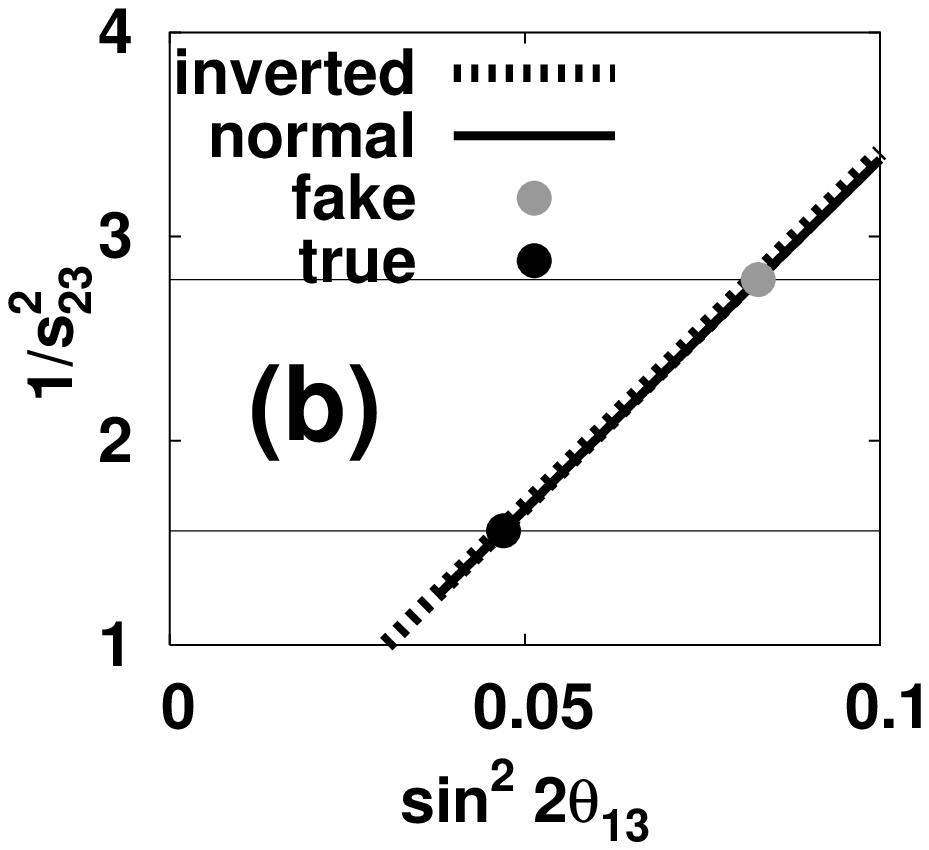}
\includegraphics[scale=0.37]{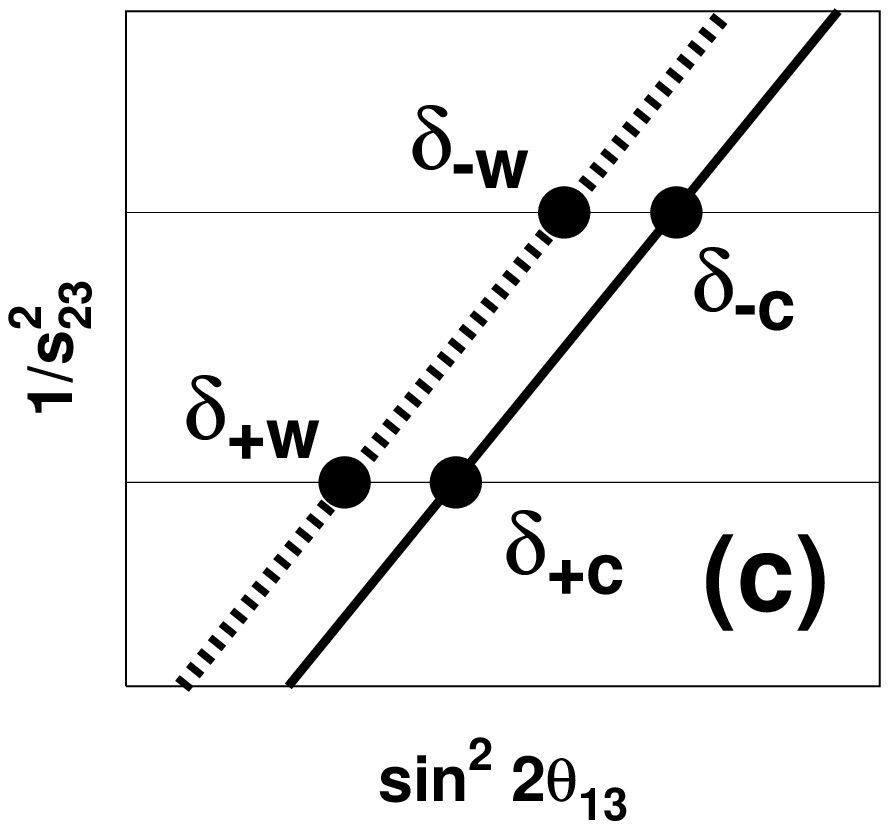}
\caption{\scriptsize
The $\theta_{23}$ ambiguity which could
arise after the JPARC measurements of
$P(\nu_\mu\rightarrow\nu_e)$,
$P(\bar{\nu}_\mu\rightarrow\bar{\nu}_e)$ and
$P(\nu_\mu\rightarrow\nu_\mu)$ at the oscillation
maximum.
(a) If $\sin^22\theta_{23}\simeq1.0$ then
the values of $\theta_{13}$ and $\theta_{23}$ are
close to each other for all the solutions.
(b) If $\sin^22\theta_{23}<1$ then
the $\theta_{23}$ ambiguity has to be resolved to determine
$\theta_{13}$ and $\theta_{23}$ to good precision.
(c) Enlarged figure of (b) with
four possible values for the CP phase $\delta$
at the oscillation maximum.  The solid (dashed) line stands for
the normal (inverted) hierarchy.
}
\label{fig5}
\end{figure}

If $\sin^22\theta_{23}\simeq1$, then all the four solutions
are basically close to each other in the 
($\sin^22\theta_{13}$, $1/s^2_{23}$) plane, and the ambiguity
due to degeneracies are not serious as far as $\theta_{13}$ and
$\theta_{23}$ are concerned (See Fig.\,\ref{fig5}(a)).
On the other hand, if $\sin^22\theta_{23}$ deviates fairly from 1,
then the solutions are separated into two groups, those for
$\theta_{23}>\pi/4$ and those for $\theta_{23}<\pi/4$
in the ($\sin^22\theta_{13}$, $1/s^2_{23}$) plane, as is shown
in Fig.\,\ref{fig5}(b).  In this case resolution of the
$\theta_{23}\leftrightarrow\pi/2-\theta_{23}$ ambiguity is necessary
to determine $\theta_{13}$, $\theta_{23}$ and $\delta$.

Resolution of the $\theta_{23}$ ambiguity has been discussed
by several groups
using the disappearance measurement of
$P(\bar\nu_e \to \bar\nu_e)$ at reactors
or the silver channel $\nu_e \to \nu_\tau$ at
neutrino factories.  Here I will discuss the
prospects of the channels $\nu_\mu \to \nu_e$,
$\bar{\nu}_\mu\rightarrow\bar{\nu}_e$ and
$\nu_e\rightarrow\nu_\tau$.

\subsection{$\nu_\mu\rightarrow\nu_e$}
From the measurements of $P(\nu_\mu \to \nu_e)$ and
$P(\bar\nu_\mu \to \bar\nu_e)$ by JPARC at
the oscillation maximum the
value of $\delta$ can be deduced up to the eightfold ambiguity
($\delta\leftrightarrow\pi-\delta$,
$\theta_{23}\leftrightarrow\pi/2-\theta_{23}$,
$\Delta m^2_{31}\leftrightarrow-\Delta m^2_{31}$).
As is depicted in Fig.\,\ref{fig5}(c), depending on whether
$s^2_{23}-1/2$ is positive or negative, I assign the
subscript $\pm$, and depending on whether our ansatz
for sgn($\Delta m^2_{31}$) is correct or wrong, I assign
the subscript c or w.  Thus the four possible
values of $\delta$ for each assumption on the mass hierarchy
are given by
\begin{eqnarray}
(\delta_{+\text{c}}, \delta_{-\text{c}},
 \pi-\delta_{+\text{c}}, \pi-\delta_{-\text{c}});~~
(\delta_{+\text{w}}, \delta_{-\text{w}},
\pi-\delta_{+\text{w}}, \pi-\delta_{-\text{w}}).
\label{8delta}
\end{eqnarray}
Now suppose that the third measurement gives the value $P$
for the oscillation probability $P(\nu_\mu \to \nu_e)$.
Then there are in general eight lines
in the ($X\equiv\sin^22\theta_{13}$, $Y\equiv 1/s^2_{23}$) plane
given by
\begin{eqnarray}
f^2X&=&\left[P-C+2C\cos^2(\delta+\Delta)\right](Y-1)+P-2\cos(\delta+\Delta)
\nonumber\\
&{\ }&\times\sqrt{C(Y-1)}
\sqrt{\left[P-C\sin^2(\delta+\Delta)\right](Y-1)+P}
\label{p3n}
\end{eqnarray}
for the normal hierarchy, and
\begin{eqnarray}
\bar{f}^2X&=&\left[P-C+2C\cos^2(\delta-\Delta)\right](Y-1)+P
-2\cos(\delta-\Delta)
\nonumber\\
&{\ }&\times\sqrt{C(Y-1)}
\sqrt{\left[P-C\sin^2(\delta-\Delta)\right](Y-1)+P}
\label{p3i}
\end{eqnarray}
for the inverted hierarchy, where
$\Delta\equiv|\Delta m^2_{31}|L/4E$ is defined for the third measurement,
and $\delta$ takes one of the four values for each assumption
on the mass hierarchy given in Eq. (\ref{8delta}).

Let me look at three typical cases:
$L$=295km, $L$=730km, $L$=3000km.
The reference values for the
oscillation parameters used here are
\begin{eqnarray}
&{\ }&\sin^22\theta_{12}=0.8,~~
\sin^22\theta_{13}=0.05,~~
\sin^22\theta_{23}=0.96,\nonumber\\
&{\ }&\Delta m^2_{21}=7\times10^{-5}\mbox{\rm eV}^2,~~
\Delta m^2_{31}=2.5\times10^{-3}\mbox{\rm eV}^2>0,~~
\delta=\pi/4,
\label{ref}
\end{eqnarray}
where I am assuming the normal hierarchy for simplicity.
\begin{figure}
\vglue -0.5cm
\includegraphics[width=3.5cm]{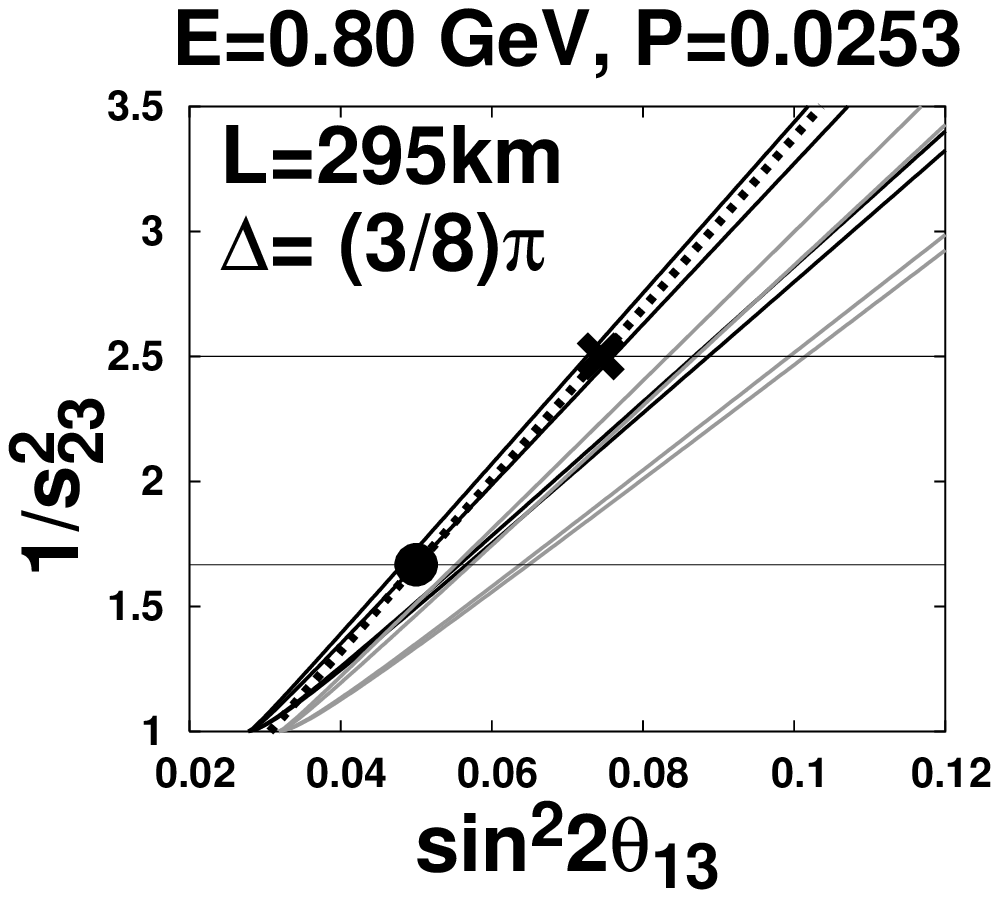}
\includegraphics[width=3.5cm]{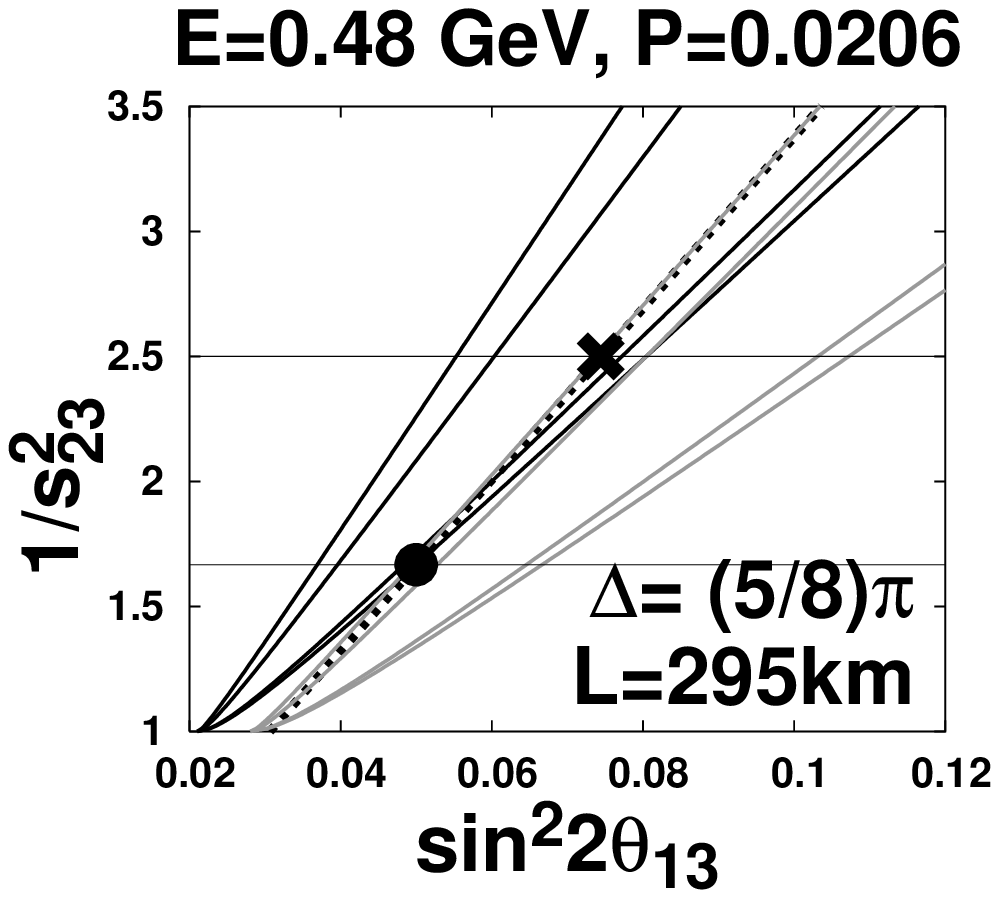}
\includegraphics[width=3.5cm]{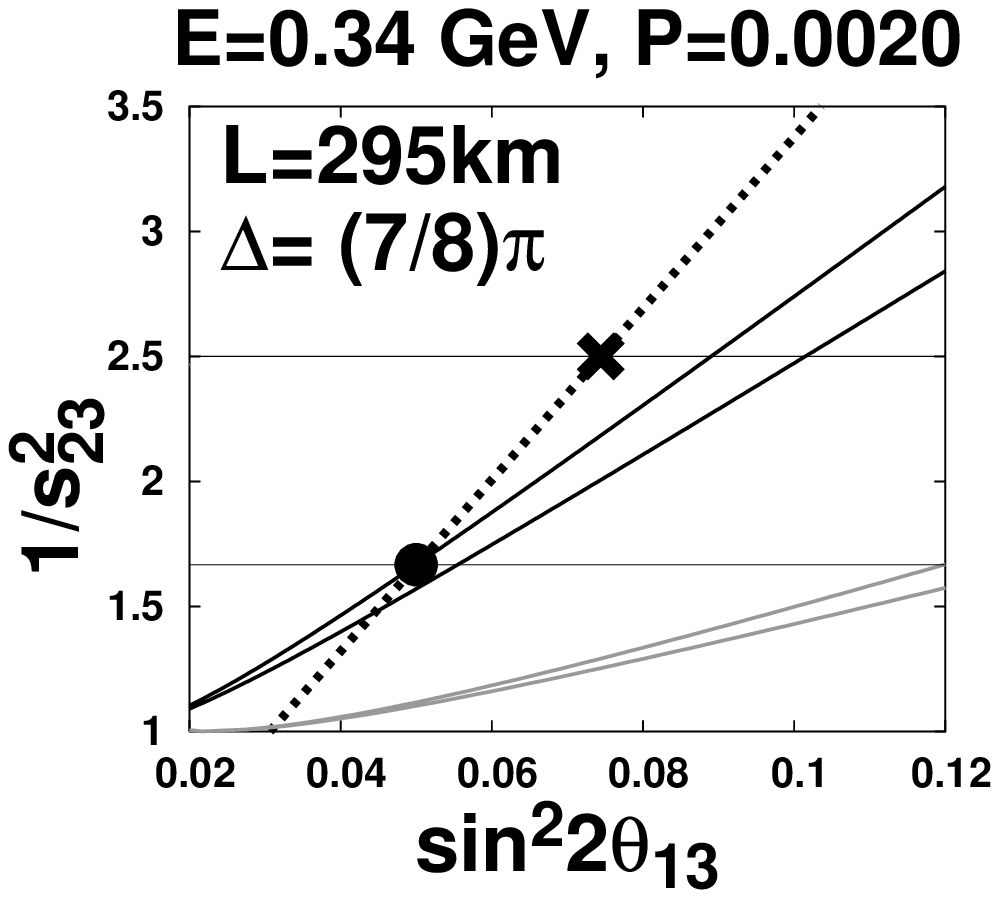}
\includegraphics[width=3.5cm]{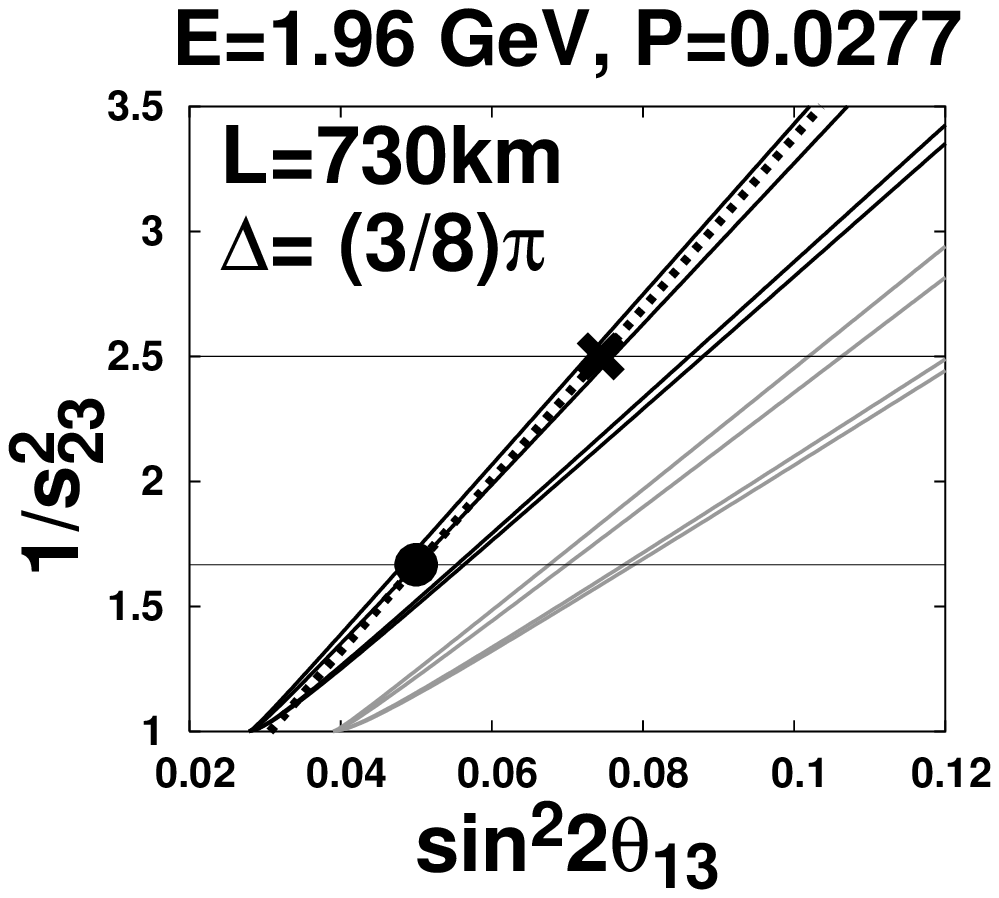}
\includegraphics[width=3.5cm]{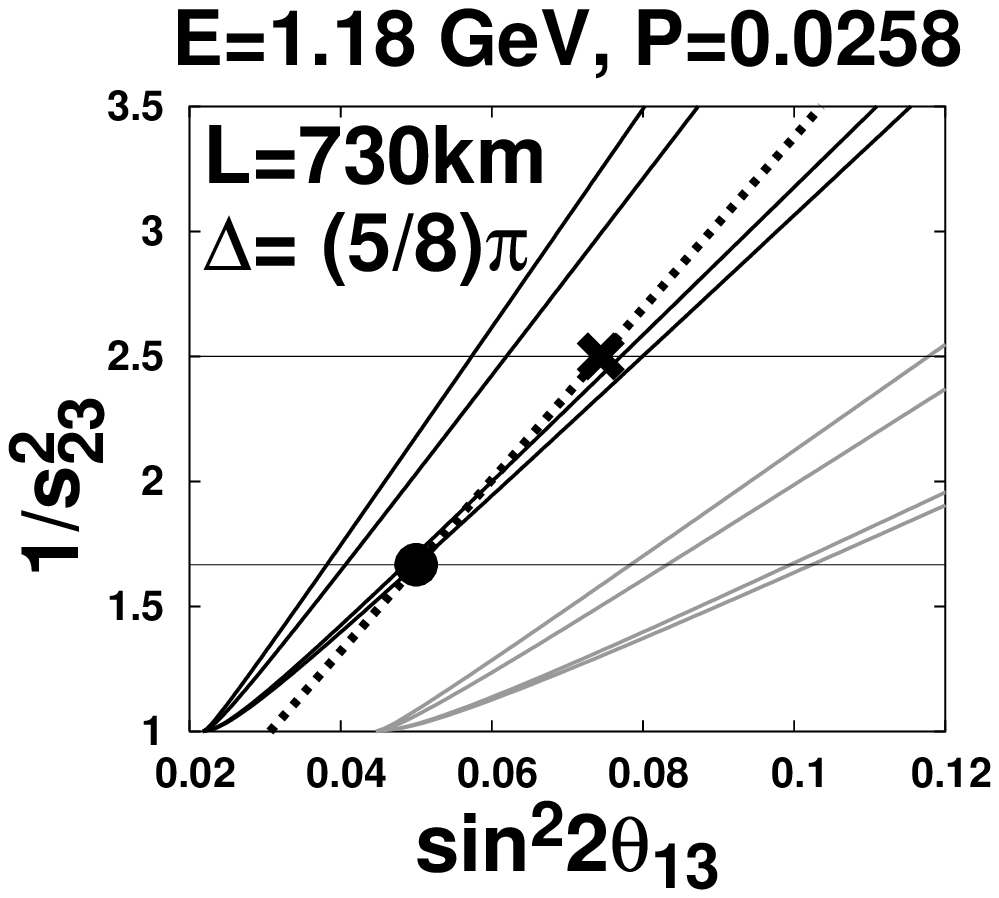}
\includegraphics[width=3.5cm]{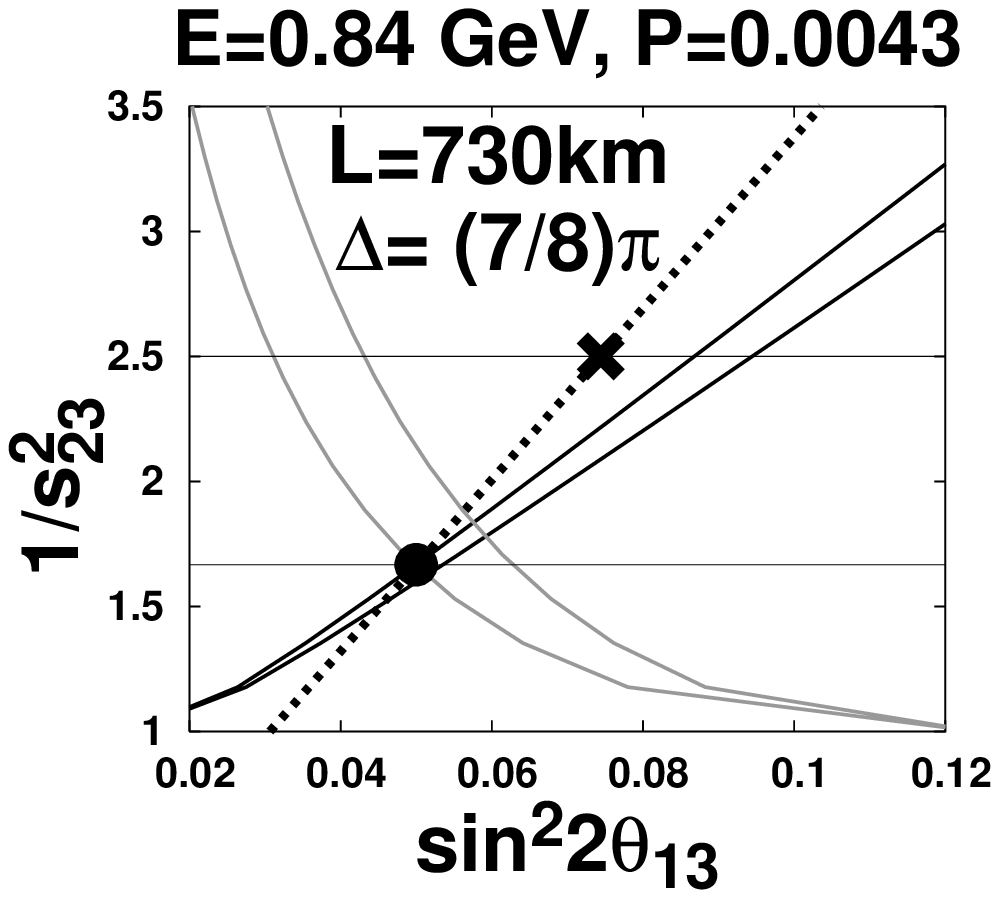}
\includegraphics[width=3.5cm]{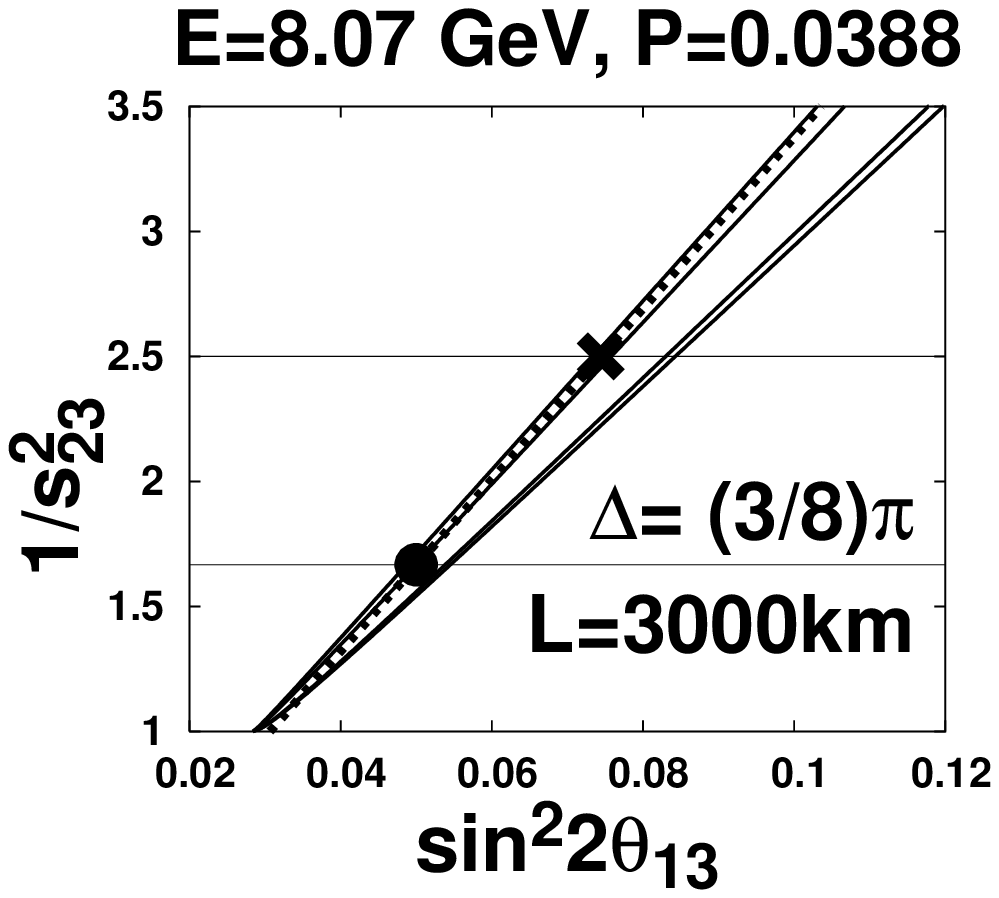}
\includegraphics[width=3.5cm]{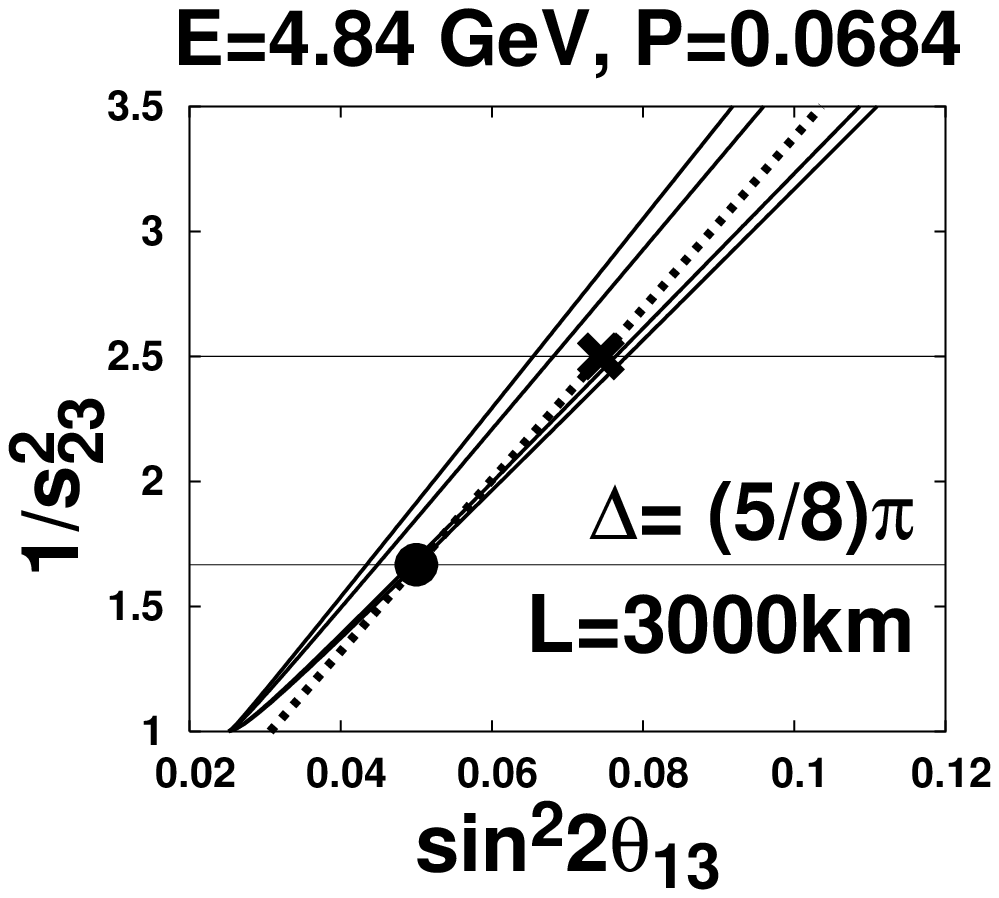}
\includegraphics[width=3.5cm]{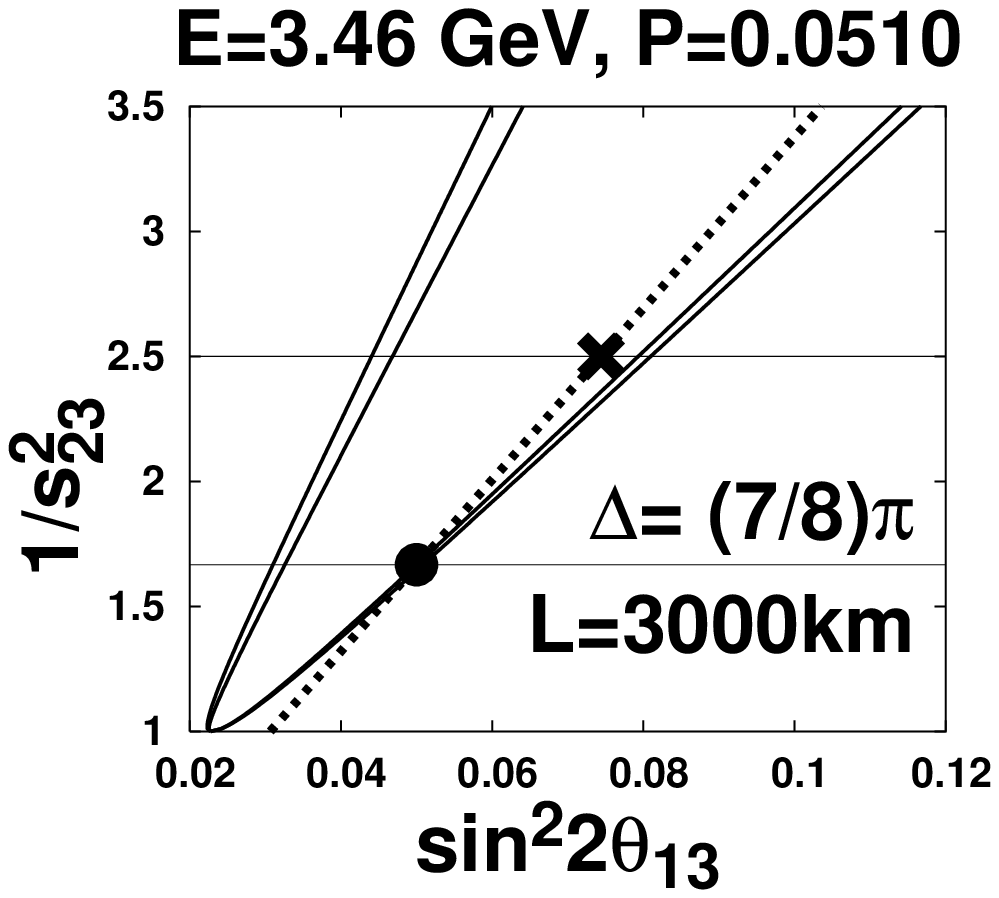}
\caption{\scriptsize
The trajectories of $P(\nu_\mu\rightarrow\nu_e)=$ const.
of the third experiment at $L$=295km, $L$=730km, $L$=3000km
with $\Delta\equiv|\Delta m^2_{31}|L/4E=(j/8)\pi~(j=3,5,7)$
after JPARC.  The true values are those in Eq. (\ref{ref}).
The dashed line is the JPARC result
obtained by $P(\nu_\mu\rightarrow\nu_e)$ and
$P(\bar{\nu}_\mu\rightarrow\bar{\nu}_e)$ at the oscillation
maximum.  The black (grey) solid lines are the trajectories
of $P(\nu_\mu\rightarrow\nu_e)$ given by the third experiment
assuming the normal (inverted) hierarchy, where $\delta$ takes
four values for each mass hierarchy.
The blob (cross) stands for the
true (fake) solution given by the JPARC results on
$P(\nu_\mu \to \nu_e)$,
$P(\bar\nu_\mu \to \bar\nu_e)$ and $P(\nu_\mu \to \nu_\mu)$.
}
\label{fig7}
\end{figure}
Fig.\,\ref{fig7} shows
the trajectories of $P(\nu_\mu \to \nu_e)$
obtained in the third measurement together with
the constraint of $P(\nu_\mu \to \nu_e)$,
$P(\bar\nu_\mu \to \bar\nu_e)$ and $P(\nu_\mu \to \nu_\mu)$
by JPARC, for $L$=295km, $L$=730km, $L$=3000km, respectively,
where $\Delta$ takes the values
$\Delta=j\pi/8$ ($j=3,5,7$).

From Eqs. (\ref{p3n}) and (\ref{p3i}) we see that the only
difference of the solutions with $\delta$ and with $\pi-\delta$
appears in $\cos(\delta\pm\Delta)$ or $\sin(\delta\pm\Delta)$.
It turns out that in order to resolve the
$\delta\leftrightarrow\pi-\delta$ ambiguity,
it is necessary to perform an experiment at $\Delta$ which
is far away from $\pi/2$.

To resolve the
$\Delta m^2_{31}\leftrightarrow-\Delta m^2_{31}$ ambiguity,
it is necessary to have a long baseline, as one can easily imagine.
What is not trivial to see is that the split of the curves with the
different mass hierarchies is larger for lower energy.
This can be seen by showing that the ratio of the
X-intercept at $Y=1$ for the normal hierarchy to that
for the inverted one deviates from one more for larger value of
$\Delta$ as long as $\Delta<\pi$.

As for resolution of the
$\theta_{23}\leftrightarrow\pi/2-\theta_{23}$ ambiguity,
it turns out that the term 
$|\cos(\delta+\Delta)|/f$ has to be small to resolve it.
This is because in order for the third measurement curve
which goes through the true point
to stay away from the fake point, the $X$-intercept at $Y=1$ of this
curve has to be far away from that of the JPARC line, and
the difference in the $X$-intercepts $X_\text{JPARC}$ and
$X_\text{3rd}$ at $Y=1$ for the JPARC line and the third measurement line
is proportional to $|\cos(\delta+\Delta)|/f$.
When $AL$ is small, in order for $f$ to be small,
$\left|\Delta-\pi\right|$ has to be small.
Furthermore, $|\cos(\delta+\Delta)|$ has to be large.
In real experiments, however, nobody knows the value of the true $\delta$
in advance,
so it is difficult to design a long baseline experiment to
resolve the $\theta_{23}\leftrightarrow\pi/2-\theta_{23}$ ambiguity.
If $\delta$ turns out to satisfy $|\cos(\delta+\Delta)|\sim1$
in the result of the third experiment, then we may be able to resolve
the $\theta_{23}$ ambiguity as a byproduct.
\vglue -0.2cm
\begin{figure}
\includegraphics[width=3.7cm]{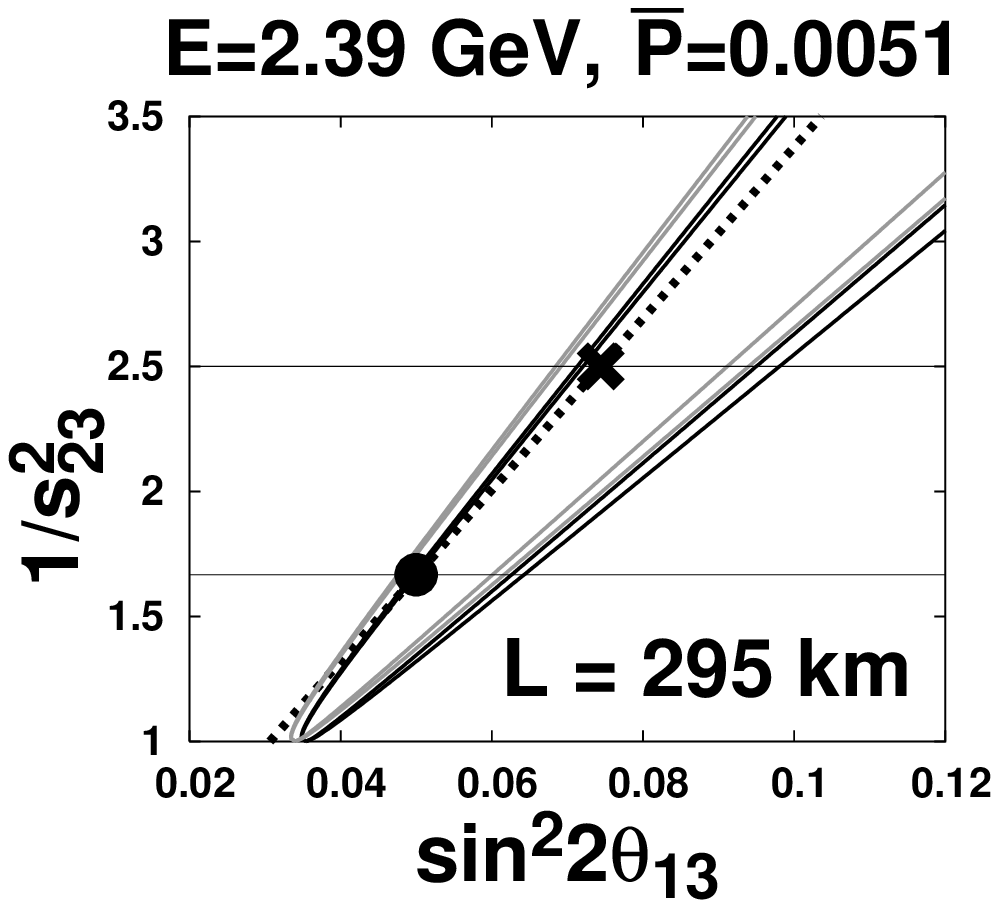}
\includegraphics[width=3.7cm]{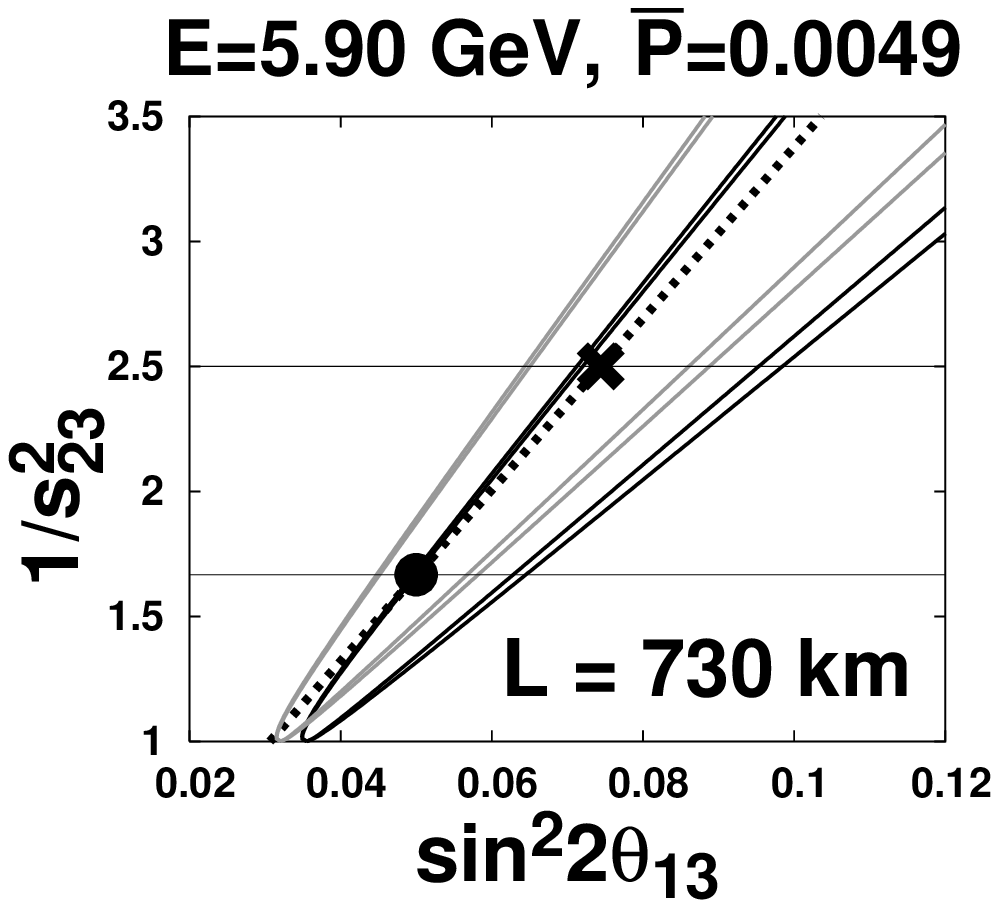}
\includegraphics[width=3.7cm]{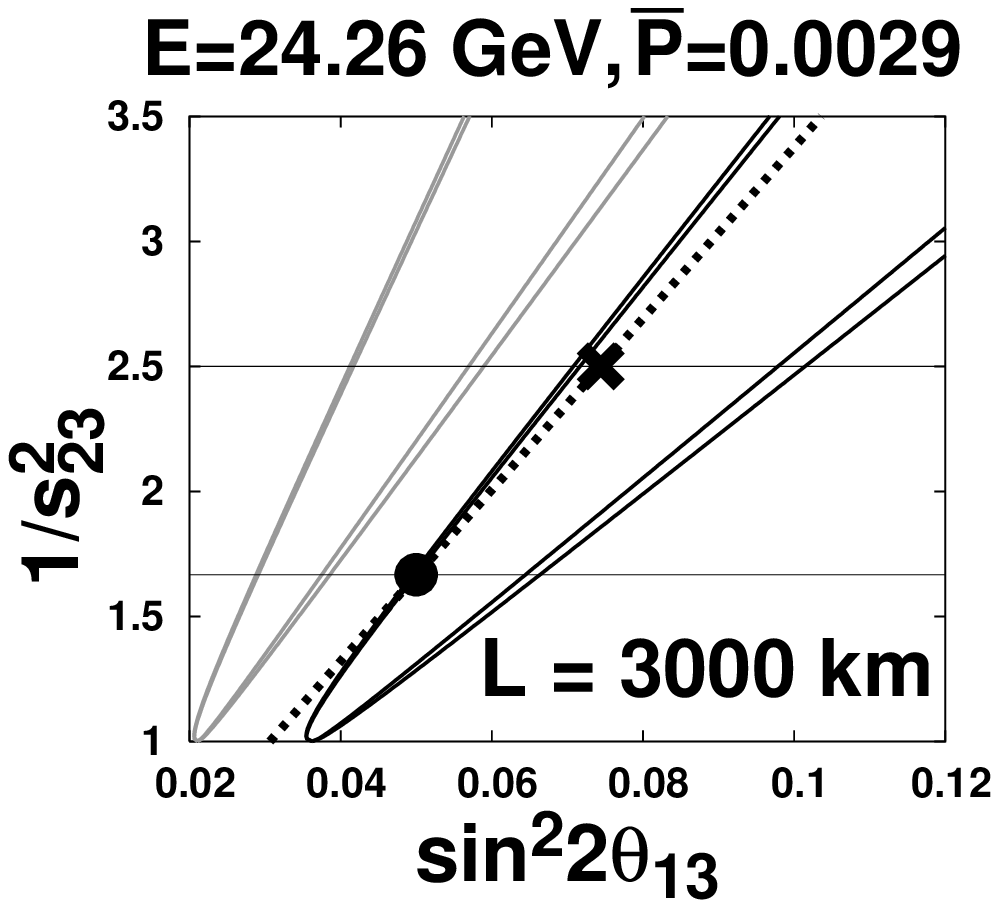}
\caption{\scriptsize
The trajectories of $P(\bar{\nu}_\mu\rightarrow\bar{\nu}_e)=\bar{P}=$ const.
of the third experiment with
$\Delta\equiv|\Delta m^2_{31}|L/4E=\pi/8$ after JPARC.
The behaviors are almost similar to those for
$P(\nu_\mu\rightarrow\nu_e)=$ const.
The true values are those in Eq. (\ref{ref}).
The blob (cross) stands for the
true (fake) solution as in Fig.\ref{fig7}}
\label{fig10}
\end{figure}
\begin{figure}
\includegraphics[width=3.7cm]{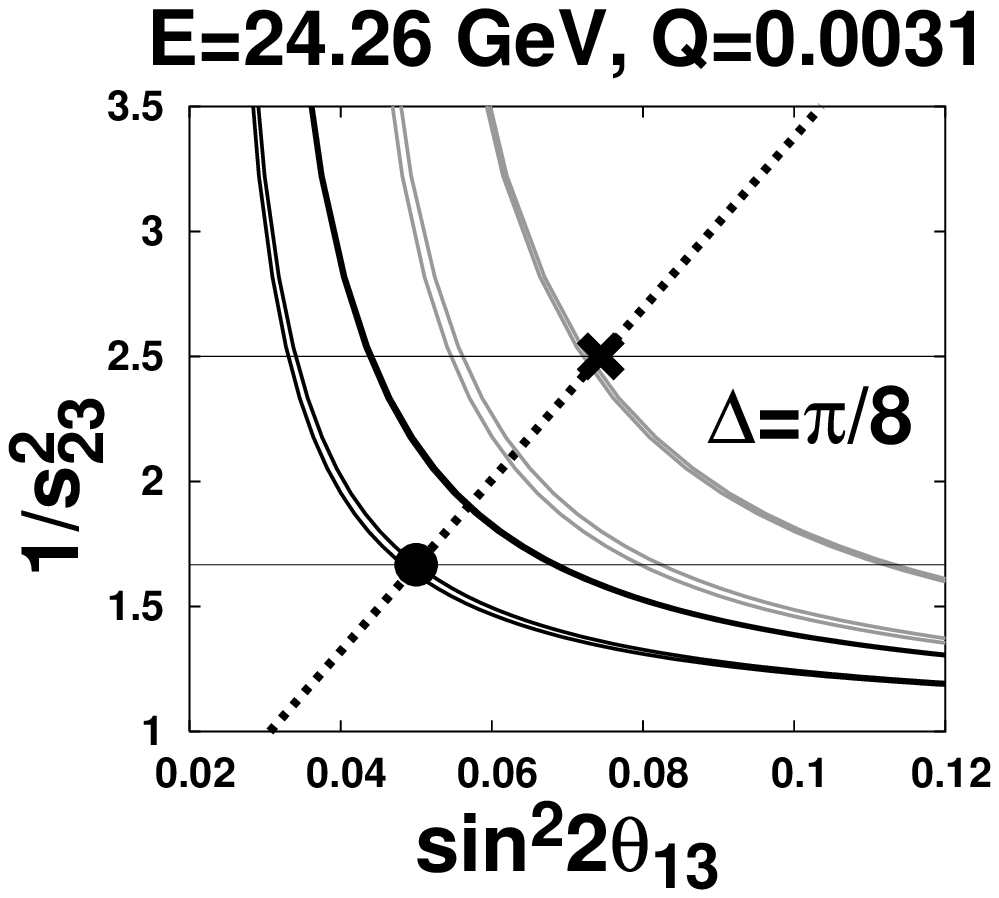}
\includegraphics[width=3.7cm]{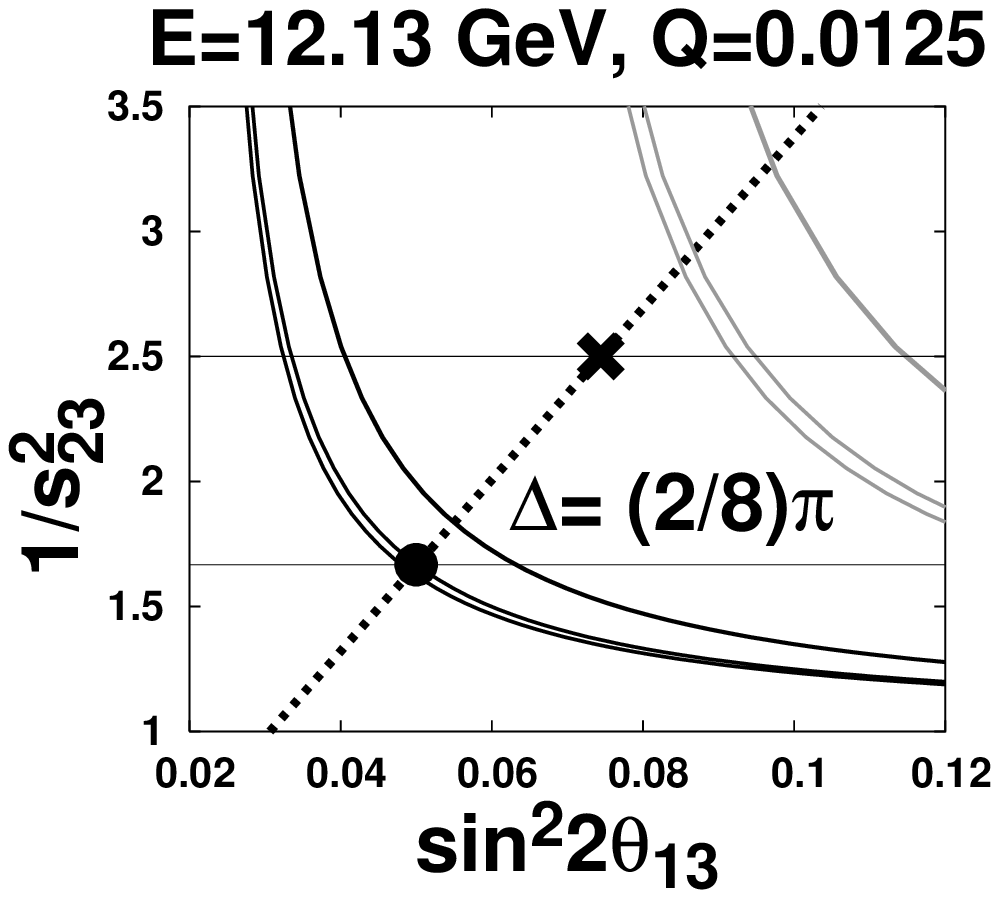}
\includegraphics[width=3.7cm]{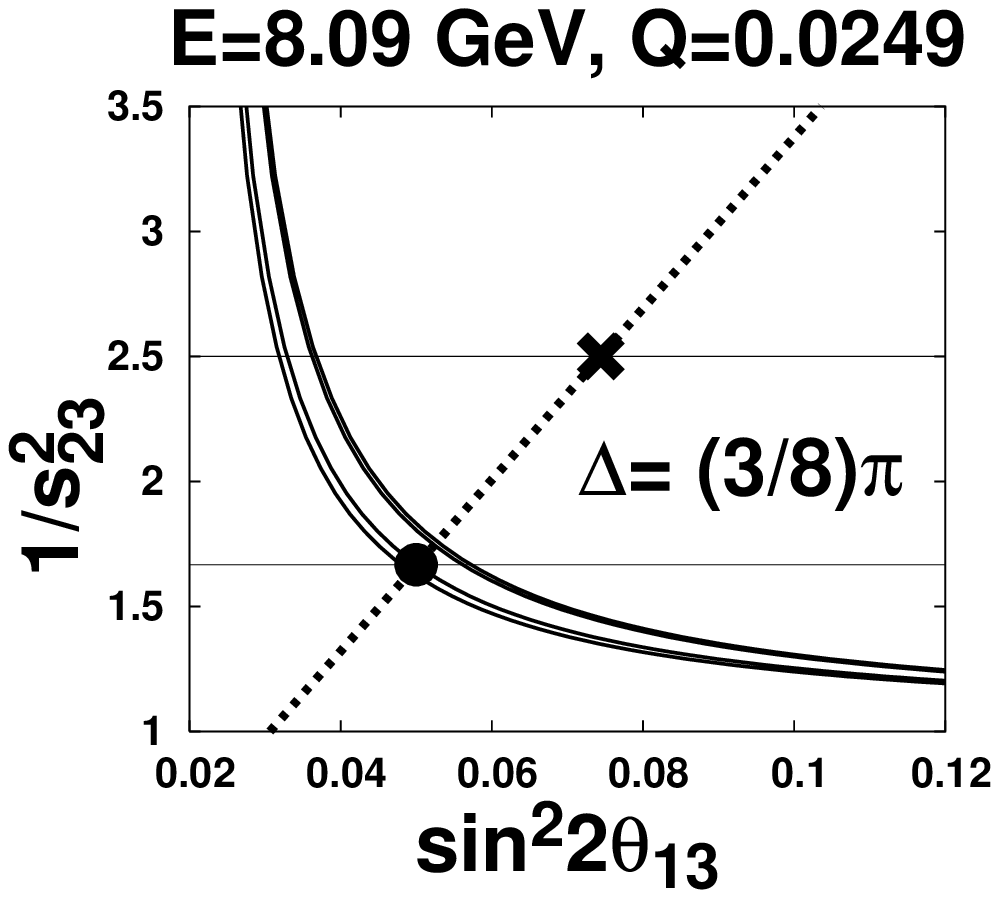}
\caption{\scriptsize
The trajectories of $P(\nu_e\rightarrow\nu_\tau)=Q=$ const.
of the third experiment at $L$=2810km
with $\Delta\equiv|\Delta m^2_{31}|L/4E=(j/8)\pi~(j=1,2,3)$
after JPARC.  The true values are those in Eq. (\ref{ref}).
The blob (cross) stands for the
true (fake) solution as in Fig.\ref{fig7}}
\label{fig11}
\end{figure}

\subsection{$\bar{\nu}_\mu\rightarrow\bar{\nu}_e$}
It turns out that the situation does not change very much even if
I use the $\bar{\nu}_\mu \rightarrow \bar{\nu}_e$ channel
in the third experiment.
Typical curves are given for $\bar{\nu}_\mu \rightarrow \bar{\nu}_e$
in Fig.\,\ref{fig10}, which are similar to those in
Fig.\,\ref{fig7}.  Thus the conclusions
drawn on resolution of the ambiguities hold qualitatively
in the case of $\bar{\nu}_\mu \rightarrow \bar{\nu}_e$ channel.

\subsection{$\nu_e\rightarrow\nu_\tau$}
The trajectory of $P(\nu_e\rightarrow\nu_\tau)=Q$,
where $Q$ is constant, in the
($X\equiv\sin^22\theta_{13}$, $Y\equiv 1/s^2_{23}$) plane
is given by
\begin{eqnarray}
X&=&\frac{Q}{f^2}\left\{\left[
1+\frac{2\cos^2(\delta+\Delta)}{1-C/Q}\right]
\frac{1-C/Q}{Y-1}+1\right.\nonumber\\
&{\ }&-\left.\frac{2\cos(\delta+\Delta)}{\sqrt{1-C/Q}}
\sqrt{\left[1+\frac{\cos^2(\delta+\Delta)}{1-C/Q}\right]
\frac{1-C/Q}{Y-1}+1}\right\}.
\label{silver}
\end{eqnarray}
Eq. (\ref{silver}) is plotted in Fig.\,\ref{fig11} in the case of
$L$=2810km.
From Fig.\,\ref{fig11} we see that the curve
$P(\nu_e\rightarrow\nu_\tau)=Q$ intersects with the JPARC
dashed line almost perpendicularly and it is experimentally advantageous:
Since the lines become thick
due to the experimental errors in reality, the allowed region is a small area
around the true solution in the $(\sin^22\theta_{13}, 1/s^2_{23})$
plane, so that one expects that the fake solution with respect to
the $\theta_{23}$ ambiguity can be excluded.
This is in contrast to the case of the $\nu_\mu\rightarrow\nu_e$
and $\bar{\nu}_\mu\rightarrow\bar{\nu}_e$ channels,
in which the slope of the black curves is almost the same
as that of the JPARC dashed line and the allowed region
can easily contain both the true and fake solutions,
so that it becomes difficult to distinguish the true point
from the fake one.
As in the case of the $\nu_\mu\rightarrow\nu_e$ channel,
the $\delta\leftrightarrow\pi-\delta$
ambiguity is expected to be resolved more likely
for the larger value of $|\Delta-\pi/2|$,
and the sgn($\Delta m^2_{31}$) ambiguity is resolved
easily for larger baseline $L$ (e.g., $L\sim$3000km).
Thus the measurement of the $\nu_e\rightarrow\nu_\tau$
channel is a promising possibility as a potentially powerful candidate
to resolve parameter degeneracies in the future.

\section{arg($U_{e3}$)}
\subsection{Fake effects on CP violation due to the
sgn($\Delta m^2_{31}$) ambiguity}
\begin{figure}
\vglue -0.3cm
\hglue -0.2cm
\includegraphics[scale=0.45]{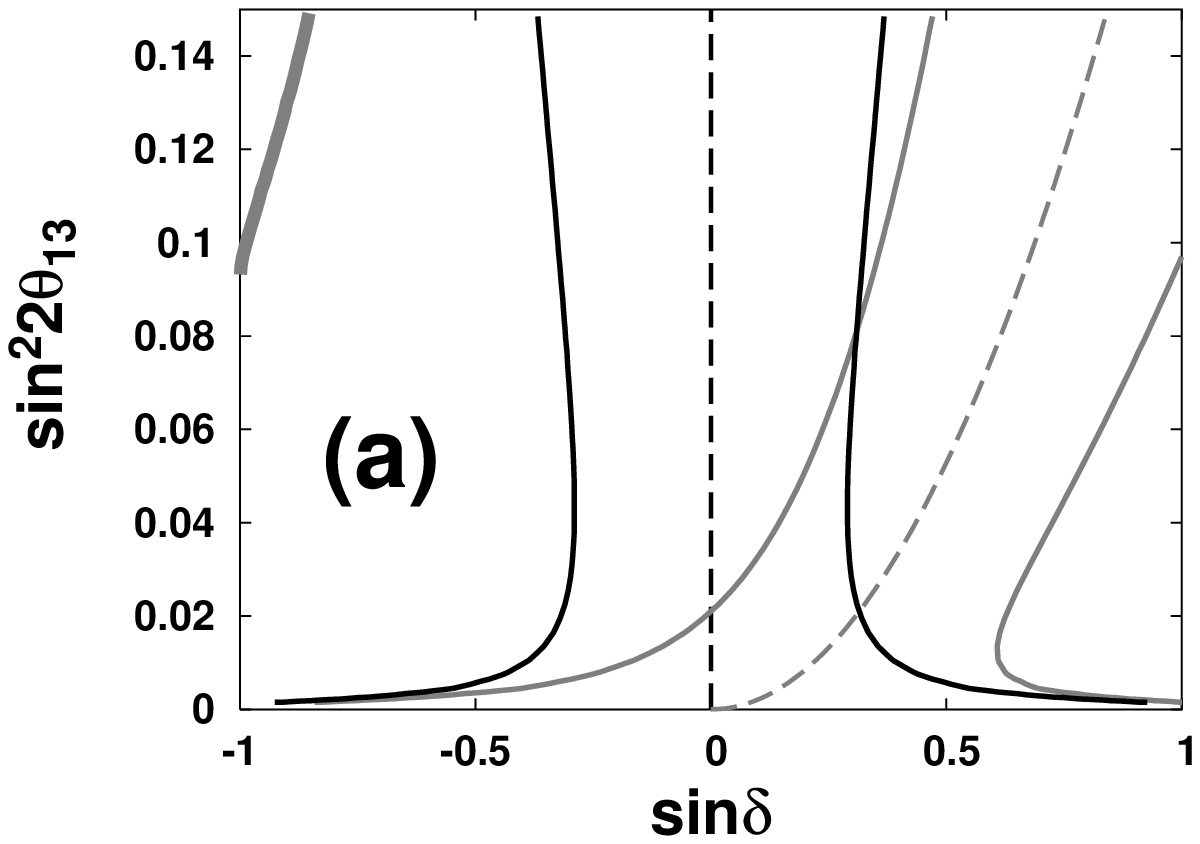}
\vglue -4.4cm\hglue 5.7cm
\includegraphics[scale=0.45]{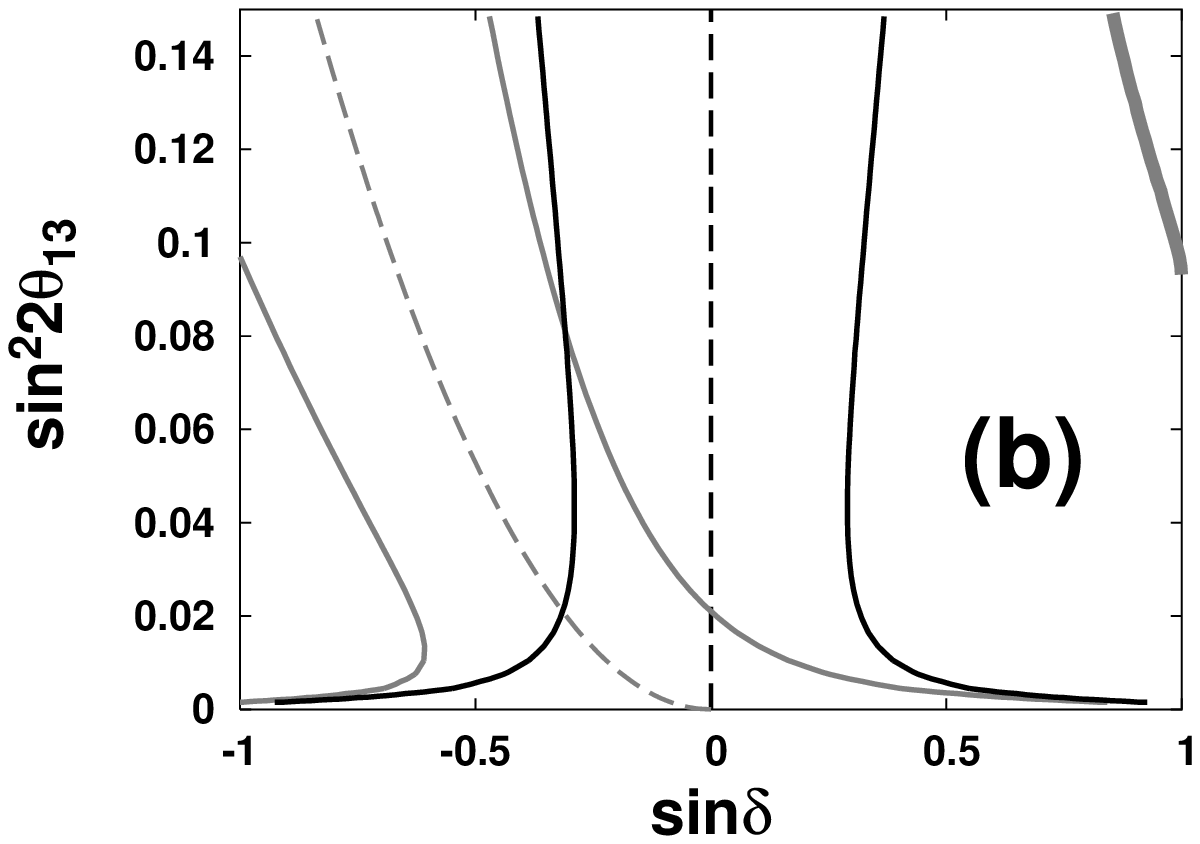}
\caption{\scriptsize The sensitivity to CP violation
at 3$\sigma$ in the case of the JPARC experiment
by a semi-quantitative analysis.  The black (grey) curves give
the sensitivity to CP violation with the correct (wrong)
assumption on the mass hierarchy when $\Delta m^2_{31}>0$
is assumed (a) or when $\Delta m^2_{31}<0$ is assumed (b).
For both black and grey curves, JPARC cannot claim for
CP violation to be nonzero inside of the area bounded by the solid curves.
These curves are plotted assuming the same width as the one obtained
in Ref.$^2$
around the center lines which are denoted by the dashed ones.
These center lines are given by $\sin\delta=0$ for the correct assumption, and
by $\sin\delta=\pm2.2\sin2\theta_{13}$ for the wrong
assumption.  The thick grey lines stand for
the boundary for the expected sensitivity at 3$\sigma$ in the
case of the combination of the JPARC $\nu_\mu\rightarrow\nu_e$
and the reactor experiments$^3$ 
for the {\it wrong} assumption
on the mass hierarchy.  The width from the center lines
in this case is so large that the curves for the correct assumption
are out of the range of this figure.
}
\label{fig1}
\end{figure}

If the true value is $\delta=0$, then the fake value $\delta'$
with respect to the sgn($\Delta m^2_{31}$) ambiguity
in the case of the JPARC experiment is given by
\begin{eqnarray}
\sin\delta^\prime \simeq-2.2\sin2\theta_{13},
\nonumber
\end{eqnarray}
which is not negligible unless $\sin^22\theta_{13}\ll 10^{-2}$.
In Fig.\ref{fig1} the region is depicted
in the ($\sin\delta$, $\sin^22\theta_{13}$) plane
in which CP violation cannot be claimed to be nonzero
in the case with the correct (wrong) assumption on the
mass hierarchy.
Therefore, to determine the CP phase to good precision,
it is important to know the sign of $\Delta m^2_{31}$.

\subsection{Fake effects on CP violation due to the
$\theta_{23}$ ambiguity}
If the true value $\delta$ is zero, then the CP phase $\delta'$
for the fake solution with respect to the $\theta_{23}$ ambiguity
in the case of JPARC is given by
\begin{eqnarray}
\left|\sin\delta'\right|\sim\frac{1}{200}
\frac{|\cot2\theta_{23}|}{t_{23}}
\frac{1}{\sin2\theta_{13}}\lesssim
\frac{1}{500}\frac{1}{\sqrt{\sin^22\theta_{13}}},
\nonumber
\end{eqnarray}
where I have used the bound $0.90\le\sin^22\theta_{23}\le1.0$
from the atmospheric neutrino data in the second inequality,
so that we see that the ambiguity due to the $\theta_{23}$
does not cause a serious problem on determination of $\delta$
for $\sin^22\theta_{13}\gtrsim10^{-2}$.

\section{Summary}
The two main conclusions are:
(1) To determine $\theta_{13}$, it is important to
resolve the $\theta_{23}$ ambiguity if $\sin^22\theta_{23}$
turns out to deviate fairly from 1;
(2) To determine $\delta$, it is important to
resolve the sgn($\Delta m_{31}^2$) ambiguity.
The possibility to resolve the $\theta_{23}$ ambiguity
was discussed in the case of $\nu_\mu\rightarrow\nu_e$,
$\bar{\nu}_\mu\rightarrow\bar{\nu}_e$ and
$\nu_e\rightarrow\nu_\tau$, using the plot
of constant probabilities in the
($\sin^22\theta_{13}$, $1/s^2_{23}$) plane.
The $\nu_e\rightarrow\nu_\tau$
channel seems to be most promising, while other two channels
$\nu_\mu\rightarrow\nu_e$ and $\bar{\nu}_\mu\rightarrow\bar{\nu}_e$ 
may be useful to resolve the $\theta_{23}$ and
$\delta\leftrightarrow\pi-\delta$ ambiguities
for $\pi/2<\Delta\equiv|\Delta m_{31}|L/4E<\pi$.
Experiments with longer baselines
($\gtrsim$1000km) are expected to determine sgn($\Delta m_{31}^2$).

\section*{Acknowledgments}
I would like to thank Hiroaki Sugiyama for many discussions.

\end{document}